\documentclass[twocolumn,showpacs,pre,superscriptaddress]{revtex4}
\begin{document}
\title{Null Frenet-Serret Dynamics}
\author{R. Capovilla}
\email{capo@fis.cinvestav.mx}
\affiliation{Departamento de F\'{\i}sica,
Centro de Investigaci\'on y de Estudios
Avanzados del IPN,
 Apdo. Postal 14-740,07000 M\'exico, DF,
MEXICO}
\author{J. Guven}
\email{jemal@nuclecu.unam.mx}
\affiliation{
Instituto de Ciencias Nucleares,
Universidad Nacional Aut\'onoma de M\'exico,
Apdo. Postal 70-543, 04510 M\'exico, DF,
MEXICO}
\author{E. Rojas}
\email{efrojas@uv.mx}
\affiliation{Facultad de F\'{\i}sica e
Inteligencia Artificial, Universidad Veracruzana,
91000 Xalapa, Veracruz, MEXICO}
\date{\today}
\begin{abstract} 
We consider the Frenet-Serret geometry of null 
curves in a three and a four-dimensional
Minkowski background. We develop a theory of 
deformations adapted to the Frenet-Serret frame. 
We exploit it to provide a Lagrangian description of
  the dynamics of  geometric models for null curves. 
\end{abstract}


\maketitle

\vskip 3em

{\it Dedicated to Mike Ryan on his sixtieth
birthday.}

\vskip 3em

The notion of a relativistic point-like 
object, or particle, is an idealization 
that has  guided the development 
of reparametrization invariant theories, in particular, 
string theory and its membrane descendants
\cite{Polchinski,Siegel}; it has also informed our 
understanding of  general relativity as a dynamical theory
\cite{Hen.Tei:95,Kuchar}.
The massive relativistic 
particle, with an action proportional to the 
length of its worldline, represents the simplest
global geometrical quantity invariant under 
reparametrizations. The particle follows a geodesic 
of the ambient spacetime -- the worldline curvature 
vanishes. A natural extension is to consider higher order 
geometrical models for particles, described by an action 
that depends on the
curvatures of the worldline. While the initial 
motivation for their introduction  was their value as toy models 
for higher dimensional relativistic systems such as 
strings \cite{1}, it has turned out that they 
possess interesting features in their own right;  
they model spinning particles \cite{2,Nes,Nerse} and they 
are relevant to various integrable systems 
\cite{3}. The approach, traditionally adopted, is to 
go Hamiltonian, with an eye on 
canonical quantization. The 
disadvantage has been that models have been examined on a case by 
case basis with a tendency to lose sight of shared features 
(see however \cite{HFS} for a 
way to remedy this shortcoming). An alternative Lagrangian
approach, developed in Ref. \cite{ACG},  
describes the dynamics of these higher order 
geometric models for particles 
in terms of the Frenet-Serret representation 
of the worldline in Minkowski spacetime. This 
representation exploits the existence 
of a preferred parametrization for curves 
-- parametrization by arclength. 
A clear advantage of this approach is that 
the conserved quantities associated 
with the underlying Poincar\'e symmetry are described directly in terms of the
{\it geometrically significant} worldline curvatures. 

In these geometrical models 
for relativistic particles, the 
worldline is timelike, so that  spacelike 
normal vector fields can be consistently defined. 
As it stands, therefore, this framework  does 
not admit null curves, where arc-length vanishes
and the normal is tangential 
to the curve. Besides their intrinsic 
interest as essentially relativistic objects, however,
null curves are also potentially valuable in the 
construction of geometric models for light-like 
relativistic extended objects \cite{nullstrings,bal,bozi}. 
An extension of the Frenet-Serret representation 
to the case of null curves was constructed 
recently \cite{NR1}, and applied 
to the simplest geometrical model based on an action
proportional to pseudo arclength involving second derivatives (see also Refs. 
\cite{Lucas} for interesting work on the 
geometry of null curves). Subsequently, in the 
special case of a particle moving in $2+1$ 
dimensions, both the model proportional to pseudo 
arclength \cite{NR2}, and a model depending on 
the first pseudo-curvature have been considered 
\cite{NMMK,FGL}. 

In this paper, we develop a theory of deformations of  
the geometry 
of null curves adapted to a Frenet-Serret frame. We work in four
spacetime dimensions. The first variations 
of the geometrical action is  described directly in terms of 
curvatures. In the case of the  
models previously considered, the description of the 
dynamics is simplified considerably. This streamlining
makes it feasable to consider interesting four-dimensional generalizations.
We consider, in particular, a model linear 
in the first curvature in a four-dimensional 
background. We show that its dynamics can be framed in terms 
of the dynamics of a fictitious non-relativistic 
particle moving in two dimensions. We also consider a 
model linear in the second curvature in a 
four-dimensional background.

We begin by briefly summarizing the Frenet-Serret 
geometry for null curves as given in Ref. \cite{NR1}, 
to which we refer the interested reader for a 
more detailed treatment.
A curve in four-dimensional Minkowski spacetime 
is described by the embedding 
\begin{equation}
x^\mu  = X^\mu  (\lambda )\,, 
\end{equation}
with $\lambda$ an arbitrary parameter
($\mu,\nu, \dots = 0,1,2,3$); $x^\mu$ are local 
coordinates in Minkowski spacetime, and $X^\mu$ 
are the embedding functions. The tangent vector 
is $\dot{X}^\mu = d X^\mu / d \lambda $ (we denote 
with an overdot a derivative with respect to 
$\lambda$). The curve is null, it lives on the
light cone, 
\begin{equation}
\eta_{\mu\nu} \dot{X}^\mu \dot{X}^\nu = 
\dot{X} \cdot \dot{X} = 0\,,
\end{equation}
where $\eta_{\mu\nu}$ is the Minkowski metric. 
We use a signature $(+--- )$ throughout the paper.  

The infinitesimal pseudo arclength for a null 
curve can be defined as 
\begin{equation}
d \sigma =  \left( -\ddot{X} \cdot \ddot{X} 
\right)^{1/4} d\lambda\,.
\label{eq:arc}
\end{equation}
This is a natural generalization of the arc-length 
 for non-null curves $d\sigma = (- \dot{X} 
\cdot \dot{X} )^{1/2}$; arc-length itself clearly will not do for 
null curves since it vanishes.
We denote with a prime derivation with respect to 
$\sigma$. The pseudo arclength, like arclength,  is invariant under reparametrizations.

The Frenet-Serret frame adapted to this class of 
curves is given by the four spacetime vectors
$\{ e_+ , e_- , e_1 , e_2 \}$, 
where 
\begin{eqnarray}
e_+ &=& X'\,,  \nonumber
\\
 e_1 &=& e_+{}' \,, \nonumber
 \\ 
e_+{}^2 &=& e_-{}^2 = 0\,, \nonumber
\\
 e_+ \cdot e_1 &=& e_+ \cdot e_2 = 
 e_- \cdot e_1 = e_- \cdot e_2 = e_1 \cdot e_2 = 0 \,,
\nonumber \\ 
e_+ \cdot e_- &=& - e_1 
\cdot e_1 = - e_2 \cdot e_2 = 1\,. \nonumber
\end{eqnarray}
Two of the vectors are null, and two are spacelike.
We assume that the curve is sufficiently 
smooth so that this frame is well defined. 

The Frenet-Serret equations for a null curve are 
\begin{eqnarray}
e_+{}' &=& e_1 \,, \\
e_1{}' &=& \kappa_1  e_+ + e_{-} \,, \label{eq:fs2}  \\
e_{-}{}' &=& \kappa_1  e_1 + \kappa_2 e_{2}  \,, 
\label{eq:fs3}\\
e_{2}{}' &=& \kappa_2  e_+ \,, \label{eq:fs4}
\end{eqnarray}
where the two curvatures are given by
\begin{eqnarray}
\kappa_1 &=& {1\over 2}  X''' \cdot X'''\,, 
\label{eq:k1}
\\
\kappa_2 &=& \sqrt{ -   X'''' \cdot X''''
- ( X''' \cdot X''' )^2  }\,.
\label{eq:k2}
\end{eqnarray}
The null curve is characterized by 
two curvatures, 
whereas a non-null curve is described by three 
curvatures, in a four-dimensional background. The 
difference is due to the fact that the null curve 
is constrained to a light cone. In a way,  $\kappa_1$ 
can be thought as the analogue of the second
Frenet-Serret curvature (or torsion) of the 
non-null case, in the sense that $\kappa_1$ depends 
on three derivatives with respect to pseudo 
arclength. Note that null curves with constant 
curvatures $\kappa_1, \kappa_2$ are helices on 
the light cone \cite{Lucas}.
If $\kappa_2=0$, the curves lives in a 2+1 dimensional
Minkowski space.

We consider reparametrization invariant models 
for null curves whose dynamics is determined by 
an action of the form
\begin{equation}
S [ X ] = \int d\sigma \; L (\kappa_1 , \kappa_2 )\,,
\label{eq:action}
\end{equation}
where the Lagrangian $L$ is an arbitrary function 
of the two curvatures $\kappa_1 , \kappa_2 $ as 
given by Eqs. (\ref{eq:k1}), (\ref{eq:k2}). The 
first variation of this action will yield both
the equations of motion and the Noether charge. 
The latter gives the conserved quantities associated 
with the underlying Poincar\'e symmetry, linear 
and angular momentum (see {\it e.g.} \cite{ACG}). 
In the first variation of the  action, we 
consider only infinitesimal deformations that mantain
the null character of the curve. 

We make an infinitesimal deformation of the curve 
\begin{equation}
X \to X + 
\delta X\,.
\end{equation}
We can expand the deformation with respect to the 
Frenet-Serret frame as
\begin{equation}
\delta X = \epsilon_+ e_+ + \epsilon_- e_- + 
\epsilon_1 e_1 + \epsilon_2 e_2\,.
\end{equation}
This is always a convenient strategy when one is 
interested in the variation of reparametrization 
independent quantities. This is because the 
deformation along $ e_+$ is an infinitesimal 
reparametrization of the curve, so that, setting $\delta_\parallel X = \epsilon_+ e_+$, for the 
infinitesimal deformation of the infinitesimal 
pseudo arclength (\ref{eq:arc}) we have
\begin{equation}
\delta_\parallel d\sigma = \epsilon_+{}' d\sigma \,.
\end{equation}
For any worldline scalar function $f (X)$, its parallel 
deformation is given by
\begin{equation}
\delta_\parallel f = \epsilon_+ f'\,.
\end{equation} 
Therefore, for the reparametrization invariant 
geometrical model defined by the action 
(\ref{eq:action}), we have
\begin{equation}
\delta_\parallel S = \int d\sigma \left( \epsilon_+ 
L \right)'\,.
\label{eq:para}
\end{equation} 
The deformation along $e_+$ contributes only a 
boundary term. The non-trivial part of the 
deformation is given by the remainder, which 
we denote by
\begin{equation}
\delta_\perp X = \epsilon_- e_- + \epsilon_1 e_1 + 
\epsilon_2 e_2\,.
\end{equation}

As we are interested only in deformations 
that preserve the null character of the curve, 
we consider 
\begin{eqnarray}
\dot{X} \cdot \delta_\perp \dot{X} &=& 
\dot{X} \cdot {d \over d\lambda} (\delta_\perp X )
\nonumber \\ 
&=& \left( { d\sigma \over d\lambda } \right)^2 e_+ 
\cdot ( \epsilon_- e_- + \epsilon_1 e_1 + 
\epsilon_2 e_2 )' \nonumber \\
&=&
 \left( { d\sigma \over d\lambda } \right)^2 
( \epsilon_1 + \epsilon_-{}' ) \,. 
\end{eqnarray}
The condition $\delta_\perp ( \dot{X} \cdot 
\dot{X} ) = 0$ implies the constraint 
\begin{equation}
\epsilon_1 + \epsilon_-{}' = 0
\label{eq:r1}
\end{equation}
on  the 
components of the deformation.
Thus deformations that preserve null curves 
are completely specified by two independent normal variations, 
$\epsilon_1, \epsilon_2$. However, in order to
keep the deformation of the geometry local in its 
components, it is convenient to take $\epsilon_- , 
\epsilon_2 $ as the independent variations. In the 
case of a three-dimensional Minkowski background, 
there is only one independent component of the normal deformation, which we take to be $\epsilon_- $.

For the  variation of the pseudo arclength 
(\ref{eq:arc}), a straightforward calculation gives, 
using (\ref{eq:r1}), 
\begin{equation}
\delta_\perp d\sigma =  {d \sigma  \over 2}  
( - \epsilon_-{}''' +  \kappa_1{}' \, \epsilon_- 
+  \kappa_2 \, \epsilon_2 ) = \Omega \, d\sigma \,,
\label{eq:nds}
\end{equation}
where we have defined the quantity $\Omega$ 
for later convenience. For any worldline scalar $f$ it 
follows that 
\begin{equation}
\delta_\perp f' = - \Omega f'
 + (\delta_\perp f )' \,.
\end{equation} 
Applying this relation to the spacetime vector $e_+$, 
we obtain  
\begin{eqnarray}
\delta_\perp e_+  &=& {1 \over 2} \left( \epsilon_-{}''' 
-  2 \kappa_1 \epsilon_-{}' - \kappa_1{}' \epsilon_- 
+  \kappa_2 \epsilon_2 \right) e_+ 
\nonumber \\
&+& \left( - \epsilon_-{}'' 
+ \kappa_1 \epsilon_- \right) e_1 
+ \left( \epsilon_2{}' + \kappa_2 \epsilon_- 
\right) e_2\,,
\label{eq:ne+}
\end{eqnarray}
and for the vector $e_1$ a similar computation produces
\begin{eqnarray}
\delta_\perp  e_1  &=& {1 \over 2} [ \epsilon_-{}''''
- \kappa_1{}'' \epsilon_- - 3 \kappa_1{}' \epsilon_-{}' - 4 
\kappa_1 \epsilon_-{}'' 
\nonumber \\
&+& 2 ( \kappa_1{}^2 + \kappa_2{}^2 ) \epsilon_- 
+  \kappa_2{}' \epsilon_2  + 3 \kappa_2 \epsilon_2{}' ]\, e_+ \nonumber \\
&+& \left( - \epsilon_-{}'' + \kappa_1 \epsilon_- \right) e_- 
+ \left( \epsilon_2{}' + \kappa_2 \epsilon_- \right)' e_2 \,.
\label{eq:ne1}
\end{eqnarray}
The variations of the other two frame
 vectors 
$e_-, e_2 $ can be calculated along the same 
lines, but we will not need them in the following.

We are now in a position to derive the  deformation
of any geometrical quantity associated with the 
curve. Let us consider the variation of the 
first curvature $\kappa_1$. 
From Eq. (\ref{eq:fs2}), we have
\begin{equation}
\delta_\perp (e_1{}' ) =  ( \delta_\perp \kappa_1 ) e_+
+ \kappa_1 \delta_\perp e_+ + \delta_\perp e_- \,,
\end{equation}
and dotting with $e_-$, we obtain
\begin{equation}
\delta_\perp \kappa_1 = e_- \cdot \delta_\perp ( e_1{}' ) 
- \kappa_1 e_- \cdot \delta_\perp e_+\,.
\label{eq:deltak1}
\end{equation}
We can read off the second term from 
Eq. (\ref{eq:ne+}). For the first term, note that
\begin{eqnarray}
e_- \cdot \delta_\perp ( e_1{}' ) &=& - \Omega 
e_- \cdot e_1{}' + 
e_- \cdot  ( \delta_\perp e_1{} )' \nonumber \\
&=& - \Omega
\kappa_1 + 
( e_- \cdot  \delta_\perp e_1{} )' 
-  e_-{}' \cdot \delta_\perp e_1  \nonumber \\
&=& - \Omega 
\kappa_1 + 
( e_- \cdot  \delta_\perp e_1{} )' -  \kappa_2 e_2 \cdot \delta_\perp e_1 \,.
\nonumber
\end{eqnarray}
Substituting these expressions in 
Eq. (\ref{eq:deltak1}), and using Eqs. (\ref{eq:ne+}), (\ref{eq:ne1}), 
we obtain 
\begin{eqnarray}
\delta_\perp \kappa_1 &=&  {1 \over 2} [ \epsilon_-{}''''
- \kappa_1{}'' \epsilon_- - 3 \kappa_1{}' \epsilon_-{}' - 4 
\kappa_1 \epsilon_-{}'' 
\nonumber \\
&+& 2 ( \kappa_1{}^2 + \kappa_2{}^2 ) \epsilon_- 
+  \kappa_2{}' \epsilon_2  + 3 \kappa_2 \epsilon_2{}' ]' \nonumber \\
&+& \kappa_2 \left( \epsilon_2{}' + \kappa_2 \epsilon_- \right)' 
+ \kappa_1 \left( \kappa_1 \epsilon_-{}' - \kappa_2 \epsilon_2 \right)\,. 
\label{eq:c1}
\end{eqnarray}

An analogous calculation gives that the variation 
of the second curvature is
\begin{eqnarray}
\delta_\perp \kappa_2 &=&
 \left[ \left( \epsilon_2{}' + \kappa_2 \epsilon_- \right)''
- \kappa_1 \epsilon_2{}' - \kappa_2 \epsilon_-{}'' \right]'
- \kappa_2 ^{2} \epsilon_2 
\nonumber \\
&+& 
\kappa_2 \kappa_1 \epsilon_-{}' - \kappa_1 (\epsilon_2{}'
+ \kappa_2 \epsilon_- )' \,.
\label{eq:c2}
\end{eqnarray}
Note that in both expressions a large part of the 
variation is in the form of a total derivative.

The expressions we have derived  allow us to obtain 
the variation of any geometric quantity 
associated with the curve. In particular, one 
can consider geometric models defined by an action 
of the form (\ref{eq:action}).
The simplest such model is proportional to pseudo 
arclength \cite{NR1,NR2}
\begin{equation}
S [X ] = 2\alpha \int d\sigma\,.
\end{equation}
Using Eqs. (\ref{eq:para}), (\ref{eq:nds}), 
its variation is found to be 
\begin{equation}
\delta S =  \alpha 
\int d\sigma \left( \kappa_1{}' \epsilon_- 
+ \kappa_2 \epsilon_2 \right)
+ \alpha  \int d\sigma \left( - \epsilon_-{}'' 
+ 2 \epsilon_+ \right)'\,.
\end{equation}
One can immediately read off the equations of 
motion to be 
\begin{equation}
\kappa_1 = \mbox{const.}\,, \quad \quad 
\quad \kappa_2 = 0\,.
\end{equation}
The solutions are null helices constrained 
to a 2+1 dimensional linear subspace of the Minkowski spacetime. 
The total derivative in the
variation of the action, using standard
techniques (see {\it e.g.} \cite{ACG}), gives the
conserved linear and angular momentum associated 
with the underlying Poincar\'e symmetry,
\begin{eqnarray}
P &=& \alpha  \left( e_- - \kappa_1 e_+ \right)\,, \\
M^{\mu\nu} &=& P^{[\mu}  X^{\nu]} + \alpha 
e^{[\mu}_+  e^{\nu]}_1 \,.
\end{eqnarray}
Note that the linear momentum is along the null 
vectors $e_+, e_- $. In this sense we can consider 
it as {\it tangential} to the curve.
The conserved mass, or first Casimir of the 
Poincar\'e group,  is
\begin{equation}
M^2 = P^2 = - 2 \alpha^2 \kappa_1\,.
\end{equation}
In order to have a positive mass, it is necessary 
that $\kappa_1 < 0$, which implies that  $e_1{}'$ 
is spacelike, as follows from Eq. (\ref{eq:k1}). 
Therefore the constant value of $\kappa_1$ is 
related to the Casimirs of the underlying Poincar\'e 
symmetry. The Pauli-Lubanski pseudo-vector is 
\begin{equation}
S_\mu = {1\over 2} {1\over \sqrt{|M^2|}} \varepsilon_{\mu\nu\rho\sigma} 
P^\nu M^{\rho\sigma} = -{\alpha^2 \over {2\sqrt{|M^2|}}}
e_{\mu\, 2}\,,
\end{equation}
with $\varepsilon_{\mu\nu\rho\sigma}$ the 
Levi-Civita tensor density, and we use the convention
$\varepsilon_{\mu\nu\rho\sigma} e^\mu{}_+ 
e^\nu{}_- e^\rho{}_1 e^\sigma{}_2 =
+ 1$.  The spin pseudo-vector is spacelike.
The second Poincar\'e Casimir is then 
\begin{equation}
|M^2 | S^2 =  -\frac{\alpha^4 }{4} \,.
\end{equation}
Moreover, we have 
\begin{equation}
S^2 = - (1/8) \alpha^2 \kappa_1 ^{-1}\,.
\end{equation}

If we consider a 2+1 ambient Minkowski 
spacetime, besides $\kappa_2 = 0$, the 
only change is in the definition of the spin 
pseudo-vector. We have 
\begin{equation}
J_\mu = \varepsilon_{\mu\rho\sigma} M^{\rho\sigma} = 
\varepsilon_{\mu\rho\sigma} P^\rho X^\sigma
- \alpha e_{\mu\, + }\,,
\end{equation}
where now we use the convention 
$\varepsilon_{\mu\nu\rho} e^\mu{}_+ e^\nu{}_- 
e^\rho{}_1 = + 1$. Note that the non-orbital 
part of $J_\mu$ is tangential. It follows that 
the second Casimir takes the form \cite{NR2}
\begin{equation}
S = J_\mu P^\mu = - \alpha^2\,.
\end{equation}

Let us consider now a model that involves the 
first curvature. The simplest one is
linear in $\kappa_1$ \cite{NMMK,FGL}, 
\begin{equation}
S = 2\int d\sigma \left( \alpha + \beta 
\kappa_1 \right)\,.
\label{eq:ak}
\end{equation}
In the simpler case of a 2+1 dimensional Minkowski 
background, using Eqs. (\ref{eq:nds}), 
(\ref{eq:c1}), we find that the model gives the 
equation of motion
\begin{equation}
\beta \kappa_1{}''' - 
{3 \beta \over 2} (\kappa_1{}^2 )'  + 
\alpha \kappa_1{}' = 0\,,
\end{equation}
which can be integrated twice to give
\begin{equation}
{1\over 2} \beta \kappa_1{}'{}^2 - { \beta \over 2} 
\kappa_1{}^3 + {1\over 2} \alpha \kappa_1{}^2 -
\gamma_{(3)} \kappa_1 = E_{(3)}\,,
\label{eq:con}
\end{equation}
where $\gamma_{(3)}$ and $E_{(3)}$ are constants that
can be expressed in terms of the Casimirs for 
this system. At the level of the curvatures, the dynamics is described by 
the motion of a fictitious particle moving in one dimension
in a cubic potential. The system is 
clearly integrable by quadratures: $\kappa_1$ can be expressed in 
terms of elliptic integrals \cite{NMMK,FGL}. It is clear from 
Eq.(\ref{eq:con}) that there are solutions with bounded periodic $\kappa_1$.
To obtain the corresponding trajectories
requires one to integrate the curvature.

The linear and angular momentum are given by
\begin{eqnarray}
P &=& \left( - \beta \kappa_1{}'' + 
\beta \kappa_1{}^2 -  \alpha \kappa_1
\right) e_+ + \beta \kappa_1{}' e_1 
\nonumber \\
&+& \left( \alpha - \beta  \kappa_1 \right) e_-
\,,
\\
M^{\mu\nu} &=& P^{[\mu} X^{\nu]} + 2 \beta e^{[\mu}{}_- e^{\nu]}{}_1 
+ \left( \alpha + \beta  \kappa_1 \right) e^{[\mu}{}_+ e^{\nu]}{}_1 \,. \nonumber \\
&&
\end{eqnarray}
Note that the linear momentum acquires a term 
in the normal direction $e_1$. This is analogous 
to what happen to curvature-dependent models
for non-null curves (see {\it e.g.} \cite{ACG}). 
The spin pseudo-vector takes the form
\begin{equation}
J_\mu = \varepsilon_{\mu\rho \sigma} P^\rho X^\sigma +
2 \beta e_{\mu\, -} - (\alpha + \beta \kappa_1 ) 
e_{\mu \, +}\,.
\end{equation}
The two Casimirs are therefore
\begin{eqnarray}
M^2 &=& 2 \left( - \beta \kappa_1{}'' 
+ \beta \kappa_1{}^2 - \alpha
\kappa_1
\right) \left( \alpha - \beta  \kappa_1 \right) 
\nonumber \\
&-& \beta^2  (\kappa_1{}' )^2 
\\
S &=& -2 \beta  \left( \beta \kappa_1{}'' 
- {3 \beta \over 2}\kappa_1{}^2 +
\alpha
\kappa_1 \right)
- \alpha^2\,.
\end{eqnarray}
The latter expression identifies the constant 
$ \gamma_{(3)}$ as
\begin{equation}
\gamma_{(3)} = - (1/2\beta) (S + \alpha^2 )\,.
\end{equation} 
We reproduce Eq.(\ref{eq:con}) subtracting the Casimirs to 
eliminate $\kappa_1{}''$:
\begin{equation}
\beta^2 \kappa_1{}'{}^2 -
\left[ \beta^{-1} ( S +\alpha^2 ) - 
\beta \kappa_1{}^2 \right] \left( \alpha - 
\beta \kappa_1 \right) = - M^2  \,,
\end{equation} 
which identifies the constant $E_{(3)}$ that appears in
Eq. (\ref{eq:con}) as
\begin{equation}
E_{(3)} = {1 \over 2 \beta^2} ( S + \alpha^2 - \beta M^2)\,. 
\end{equation}

We extend now our consideration of this 
model to a 3+1 dimensional background.
The variation of the action  (\ref{eq:ak}) 
gives the two equations of motion
\begin{eqnarray}
\beta  \kappa_1{}''' - {3 \beta \over 2} (\kappa_1{}^2 )' 
- \beta  (\kappa_2{}^2 )' + \alpha \kappa_1{}' &=& 0\,, 
\label{eq:em1}\\
2 \beta \kappa_2{}'' - \beta  \kappa_1 \kappa_2 
+ \alpha \kappa_2 &=& 0\,.
\label{eq:em2}
\end{eqnarray}
The first equation again possesses a first integral,
\begin{equation}
\beta  \kappa_1{}'' - {3 \beta \over 2} \kappa_1{}^2  
- \beta \kappa_2{}^2  + \alpha \kappa_1 = 
\gamma_{(4)} \,,
\label{eq:con2}
\end{equation}
where $\gamma_{(4)}$ is another constant.
We have therefore two coupled differential 
equations of second order. Unlike the 2+1 case, the
presence of $\kappa_2$ stymies the second integration of 
Eq.(\ref{eq:con2}). However, 
it is clear from Eqs.(\ref{eq:em2}) and (\ref{eq:con2}) that 
they can be derived from a potential: we have
\begin{equation}
 {1\over 2}\beta  (\kappa_1{}' )^2 + 2 \beta (\kappa_2{}' )^2 + 
V(\kappa_1,\kappa_2) = E_{(4)}\,,
\label{eq:con3}
\end{equation}
where
\begin{equation}
V(\kappa_1,\kappa_2) = 
- {1 \over 2} \beta
\kappa_1{}^3 + {1\over 2} \alpha \kappa_1{}^2 -
(\gamma_{(4)} + \beta \kappa_2{}^2 )\, \kappa_1  + \alpha \kappa_2{}^2\,,
\label{eq:V}
\end{equation}
and $E_{(4)}$ is
another constant. 
The dynamics is described by the motion of a fictitious 
particle moving two dimensions. 

The linear momentum is changed by the addition 
of a term in the direction $e_2$,
\begin{eqnarray}
P &=& \left( - \beta \kappa_1{}'' 
+ \beta \kappa_1{}^2 - \alpha \kappa_1
\right) e_+ + \left( \alpha - \beta 
\kappa_1 \right) e_-
\nonumber \\
&+& \beta  \kappa_1{}' e_1 + 2 \beta 
\kappa_2{}' e_2\,,
\end{eqnarray}
so that
\begin{eqnarray}
M^2 &=& 2 \left( - \beta \kappa_1{}'' 
+ \beta \kappa_1{}^2 -  \alpha
\kappa_1
\right) \left( \alpha - \beta \kappa_1 \right) 
\nonumber \\
&-& \beta^2  (\kappa_1{}' )^2 - 4 \beta^2 
(\kappa_2{}' )^2 \,.
\end{eqnarray}
The conserved angular momentum is modified to
\begin{eqnarray}
M^{\mu\nu} &=& P^{[\mu} X^{\nu]} + 2 \beta e^{[\mu}{}_- e^{\nu]}{}_1 + \left( \alpha + \beta  \kappa_1 \right) e^{[\mu}{}_+ e^{\nu]}{}_1 
\nonumber \\ 
&+& 2 \beta \kappa_2 e^{[\mu}{}_+ e^{\nu]}{}_2\,,
\end{eqnarray}
and the spin pseudo-vector takes the form
\begin{eqnarray}
2 \sqrt{|M^2|}S &=& 2 \beta  \left(  
\beta \kappa_2 \kappa_1{}'
- \beta \kappa_1 \kappa_2{}' -
\alpha \kappa_2{}' \right) e_+ 
\nonumber \\
&+& 4 \beta^2 \kappa_2{}' e_- + 
2 \beta \kappa_2 \left( \alpha - \beta 
\kappa_1 \right) e_1 \nonumber \\ 
&+&  \left( - 2 \beta^2  \kappa_1{}'' 
+ 3 \beta^2 \kappa_1{}^2 - 2 \alpha \beta 
\kappa_1  - \alpha^2 \right) e_2\,.\nonumber \\
&&  
\end{eqnarray}
Now using the conservation law (\ref{eq:con2}), we have
\begin{eqnarray}
M^2 S^2 &=& 2 \beta^3 \kappa_2{}' \left( \beta \kappa_2 \kappa_1{}' 
- \beta \kappa_1 \kappa_2{}' -  \alpha \kappa_2{}' \right) 
\nonumber \\
&-& 
\beta^2 \kappa_2{}^2 \left( \alpha - \beta 
\kappa_1 \right)^2 
- {1 \over 4} \left( \alpha^2 + 2\beta 
\gamma_{(4)} \right)^2\,,
 \end{eqnarray}
together with
\begin{eqnarray}
M^2 &=& - 2 \left( \gamma_{(4)} + \beta \kappa_2{}^2 
+ {\beta \over 2} 
\kappa_1{}^2   \right) \left( \alpha - \beta  \kappa_1 \right) 
\nonumber \\
&-& \beta^2  (\kappa_1{}' )^2 - 4 \beta^2 (\kappa_2{}' )^2 \,.
\end{eqnarray}
The latter reproduces 
Eq.(\ref{eq:con3}) with the identification 
\begin{equation}
2 \beta E_{(4)} = - M^2 - 2 \alpha \gamma_{(4)}\,.
\end{equation}
There are two first order equations for $\kappa_1$ and $\kappa_2$.
which suggests that the system is integrable, 
We are unable, however, to find an explicit reduction.

Finally, we comment briefly on a model 
linear in the second curvature
\begin{equation}
S [X] = 2 \lambda \int d\sigma\, \kappa_2 \,.
\label{eq:Sk2}
\end{equation}
Using Eqs. (\ref{eq:nds}) and (\ref{eq:c2}), the 
variation of the action (\ref{eq:Sk2}) gives the
corresponding equations of motion
\begin{eqnarray}
\lambda \kappa_2 ''' - 2\lambda \kappa_1 \kappa_2 '
+ \lambda \kappa_1 ' \kappa_2 
&=& 0 \,, \label{eq:k21}\\
2\lambda \kappa_1 '' + \lambda \kappa_2 ^{2}
&=& 0\,.\label{eq:k22}
\end{eqnarray}
These equations can be decoupled.
One solves Eq.(\ref{eq:k21}) for $\kappa_1$:
\begin{equation}
\kappa_1 = -\kappa_2{}^2 \int\, d\sigma \; \kappa_2 '''/ \kappa_2{}^3\,,
\end{equation}
and substitutes into Eq.(\ref{eq:k22}), which
gives a fifth order equation for $\kappa_2$ alone.

To summarize, the Frenet-Serret frame provides 
a natural description of a null curve.
We have shown how the deformations
of the curve can be described in a way which is adapted to the frame; 
in particular, we have considered deformations that preserve the null
character
of the curve and obtained explicit expressions for the deformations
of the curvatures.
  These  curvatures 
are used to construct 
geometrical models for null curves.
We
have examined  
the first variation of several simple  actions, demonstrating that the 
corresponding Euler-Lagrange equations can be  cast as a set of coupled 
non-linear ODEs for the curvatures.

\vskip 1cm
\noindent{\bf Acknowledgements}
\vskip .5cm
We acknowledge partial support  from CONACyT under
grants 44974-F, C01-41639 and PROMEP-2003.

\end{document}